\documentclass[]{fundam}
\usepackage{url} 
\usepackage[ruled,lined]{algorithm2e}

\usepackage{graphicx, amssymb, amsmath, tikz, booktabs, subcaption,xr}
\usetikzlibrary{arrows, decorations.markings}

\newcommand\x{0.65}
\begin{document}

%
\setcounter{page}{1}
\publyear{2021}
\papernumber{0001}
\volume{178}
\issue{1}
%

\title{Steady state distributions in generalized exclusion processes}

\address{kmgonzales1{@}up.edu.ph}

\author{Ken Joffaniel Gonzales\\
Department of Physical Sciences and Mathematics\\
University of the Philippines Manila \\ Manila, Philippines\\
kmgonzales1{@}up.edu.ph
} 

\maketitle

\runninghead{K. Gonzales}{Steady state distributions in generalized exclusion processes}

\begin{abstract}
  	The asymmetric simple exclusion process (ASEP) is a model of particle transport used in the study of biological processes such as mRNA translation. In 2014, Zhao and Krishnan introduced a new approach for analyzing the ASEP using probabilistic Boolean networks (PBN). In this paper, we revisit and further explore the PBN approach, with focus on computing steady state distributions. Explicit forms of the structure matrices of some common transitions are obtained. In addition, we derive a simplified method for computing the structure matrices of Boolean functions and a general method for writing the Boolean functions. These methods are also extended to multi-valued logic networks  for application in multi-species exclusion processes.
\end{abstract}

\begin{keywords}
Boolean network, probabilistic Boolean network, exclusion process, steady state distribution
\end{keywords}

	\section{Introduction}
\label{Introduction}

Many biological processes involve the transport of materials along a tube or track. Different approaches have been developed to analyze these processes, including the use of ordinary differential equations, numerical simulations and codon-based discrete models such as the \emph{asymmetric simple exclusion process} (ASEP). Von de Haar~\cite{Haa} gives a comparative overview of these methods. The ASEP was originally introduced by MacDonald and Gibbs~\cite{MacGib} to model the movement of ribosomes along the mRNA chain during mRNA translation. In the ASEP, the transport of materials is represented by particles hopping along a finite one-dimensional lattice. In the case of mRNA translation, the particles are the ribosomes and the sites of the lattice correspond to the codons of the mRNA. The study of the ASEP has attracted research in many areas outside biology including mathematics, physics and computer science due to its applicability in modeling other particle transport systems such as vehicular traffic~\cite{Sch} and motor-protein transport~\cite{Ner}.

Determining the steady state distribution is of particular interest in studying the ASEP and a wide array of methods are used, from matrix ansatzes~\cite{Der} to combinatorial enumeration~\cite{Woo}. Zhao and Krishnan~\cite{Zha} introduced an approach using probabilistic Boolean networks (PBN). This method  enables the computation of the transition matrix by constructing the Boolean functions that describe the transitions and computing the corresponding structure matrices of the Boolean functions.

In this paper, we revisit and further explore the PBN approach. The paper is organized as follows. In Section~\ref{ASEP}, we recall the basics of the ASEP. In Section~\ref{Boolean}, we discuss relevant details from the theory of Boolean networks, semi-tensor products and the PBN approach. We give the explicit structure matrices for Boolean functions involving classical transitions in Section~\ref{matrix}. The explicit forms of the matrices allow a more direct computation of the steady state distribution. Next, in Section~\ref{simplified} we give a simplified method for computing the structure matrices, a key step in determining the steady state distribution using the PBN approach. This simplified method does not use algebraic manipulations on semi-tensor products, and therefore saves computational run-time. Finally, in Section~\ref{general}, we give a general method for writing the logical functions. The results in Sections~\ref{simplified}~and~\ref{general} both cover Boolean and multi-valued logic networks. Unlike previous methods, they also do not require that the logical functions be written in terms entirely of logical operators, which can be a challenging task for complicated transitions.

While this paper focuses on exclusion processes, all results can be applied in general to any probabilistic Boolean and multi-valued logic networks.

\section{Exclusion processes}\label{ASEP}
The dynamics of the ASEP involve the set of allowable transitions, the rates of these transitions, the properties of the boundaries of the lattice and the update rule used. Transitions in the classical ASEP are entry, exit and hopping. For most biological processes, distinct entry and exit endpoints are assumed, usually the left and right boundaries of the lattice, respectively. Hopping rates are \emph{asymmetric}, i.e., particles have a preferred hopping direction. When only one hopping direction is allowed, typically towards the right, we get a special case of the ASEP called the \emph{totally asymmetric simple exclusion process} (TASEP). \emph{Langmuir kinetics} are introduced by allowing particles to attach or detach at site other than the entry and exit points.

The \emph{lattice} is the discrete medium through which the particles move. The classical ASEP uses a finite, one-dimensional lattice, but other finite lattice structures have also been considered (cf.~\cite{Jos}). The boundaries of the lattice are said to be \emph{open} if the endpoints are distinct and \emph{periodic} if otherwise.

The order in which the transitions are implemented depends on the update rule used, the most common of which is the \emph{random-sequential update rule}, which is performed as follows. First, the set of all possible transitions given the current state is enumerated. The next transition is determined by simulating a discrete distribution with the probabilities of the possible transitions. The state is then updated and a new set of  possible transitions is enumerated, repeating the process. For other update rules, see \cite{Sch, Raj}.

In this paper, we assume that the lattice has open boundaries and that states change according to the random-sequential update rule. The length of the lattice is denoted by $N$ and the sites of the lattice are labeled $1,2,\ldots,N$ from left to right. Figure~\ref{transallow} illustrates some common transitions. 

\setcounter{figure}{0}
\begin{figure}[h!]
	\centering{
		\scalebox{\x}{
			\begin{tikzpicture}
				[place/.style={circle,draw=blue,fill=blue,thick,inner sep=0pt,minimum size=6mm},]
				
				\foreach \x in {0,1,2,3,4,5,6,7,8,9} { \draw (\x,0) rectangle (\x+1,1);};
				
				\node (a) at (0.5,0.5) [place] {};
				\node (b) at (3.5,0.5) [place] {};
				\node (b) at (6.5,0.5) [place] {};
				\node (c) at (9.5,0.5) [place] {};
				
				\draw [->,>=stealth,line width=0.25mm] (-1,1.1) to [bend left=30] (-0.15,1);
				\draw [->,>=stealth,line width=0.25mm] (-0.15,-0.15) to [bend left=30] (-1,-0.18);
				\draw [->,>=stealth,line width=0.25mm] (3.3,1.2) to [bend left=-60] (2.5,1.2);
				\draw [->,>=stealth,line width=0.25mm] (3.7,1.2) to [bend left=60] (4.5,1.2);
				\draw [->,>=stealth,line width=0.29mm] (6.35,1.2) to [bend left=0] (6.35,1.9);
				\draw [->,>=stealth,line width=0.29mm] (6.65,1.9) to [bend left=0] (6.65,1.2);
				\draw [->,>=stealth,line width=0.25mm] (10.2,1.1) to [bend left=30] (11.15,1);
				\draw [->,>=stealth,line width=0.25mm] (11.15,-0.15) to [bend left=30] (10.2,-0.18);
				
				\node[scale=1.4] at (-0.6,1.6) {$\alpha$};
				\node[scale=1.4] at (-0.6,-0.8) {$\delta$};
				\node[scale=1.4] at (2.9,1.9) {$q$};
				\node[scale=1.4] at (4.2,1.9) {$p$};
				\node[scale=1.4] at (5.6,1.5) {$\omega_{A}$};
				\node[scale=1.4] at (7.45,1.5) {$\omega_{D}$};
				\node[scale=1.4] at (10.7,1.6) {$\beta$};
				\node[scale=1.4] at (10.7,-0.8) {$\gamma$};
				
		\end{tikzpicture}
		}
	}
	\caption{Transitions in the ASEP}
	\label{transallow}
\end{figure}
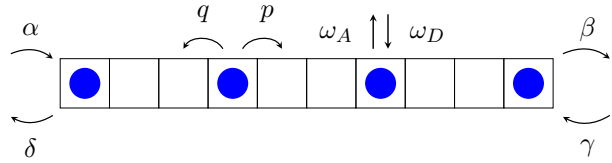

\begin{figure}
	\begin{center}
		\includegraphics[scale=0.5]{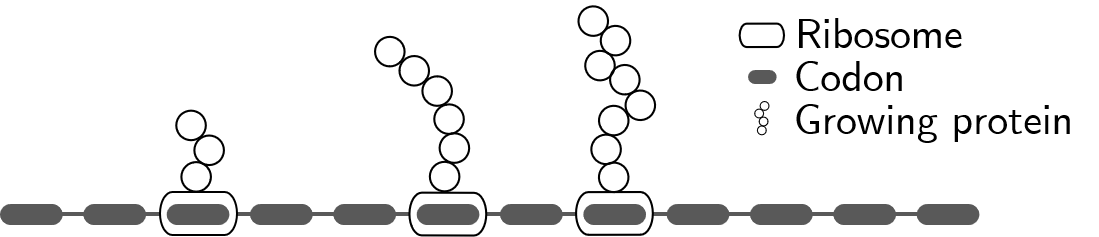}
	\end{center}
	\caption{mRNA translation}
	\label{xlation}
\end{figure}

The movement of ribosomes along the codons of the mRNA is a commonly-cited application of the ASEP in biology, specifically the TASEP. During mRNA translation, ribosomes attach on one end of the mRNA and hop along each codon towards the other end. As a ribosome travels, it assembles amino acids based on the sequence coded by the mRNA. Figure~\ref{xlation} provides a simplified illustration of this process, focusing on the movement of ribosomes. Note that other processes such as those happening within the ribosome level are not represented in the ASEP.

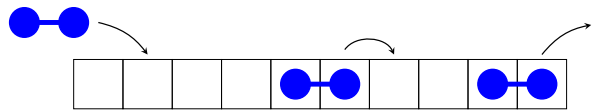
\begin{figure}
	\begin{center}
		\scalebox{\x}{
			\begin{tikzpicture}
				[place/.style={circle,draw=blue,fill=blue,thick,
					inner sep=0pt,minimum size=6mm},]
				
				\foreach \x in {0,1,2,3,4,5,6,7,8,9}
				{
					\draw (\x,0) rectangle (\x+1,1);
				};
				
				\node at (-1,1.75) [place] {};
				\node at (0,1.75) [place] {};
				\draw [-,line width=1mm,color=blue] (-1,1.75) to (0,1.75);
				
				\node at (4.5,0.5) [place] {};
				\node at (5.5,0.5) [place] {};
				\draw [-,line width=1mm,color=blue] (4.5,0.5) to (5.5,0.5);
				
				\node at (8.5,0.5) [place] {};
				\node at (9.5,0.5) [place] {};
				\draw [-,line width=1mm,color=blue] (8.5,0.5) to (9.5,0.5);
				
				\draw [->,>=stealth,line width=0.25mm] (0.5,1.75) to [bend left=20] (1.5,1.1);
				\draw [->,>=stealth,line width=0.25mm] (5.5,1.2) to [bend left=60] (6.5,1.1);
				\draw [->,>=stealth,line width=0.25mm] (9.5,1.1) to [bend left=20] (10.5,1.7);
		\end{tikzpicture}}
	\end{center}
	\caption{An ASEP model of ribosomes traveling in groups of $r$}
	\label{ZhaoModel}
\end{figure}

Most models assume that the ribosomes occupy one codon at a time. In Zhao and Krishnan's~\cite{Zha} model, the ribosomes occupy $r$ codons at a time and the entry of a ribosome happens only when the first $r$ codons are empty while the exit is initiated when the head of the ribosome is at site $N$ (see Figure~\ref{ZhaoModel}). 

The hopping rate is assumed to be the same for every particle and at every site. Tr\"osemeier et al.~\cite{Tro} introduced the codon-specific elongation model (COSEM) where a codon sequence $j$ at site $i$ is assigned the hopping rate of $\omega_{j,i}$. On the other hand, Neri~\cite{Ner} studied ASEP with bidirectional movement called the \emph{partially asymmetric exclusion process} (PASEP) over a system involving single and multiple lattices to model the transport of materials along filaments in the cellular cytoplasm.

\section{Probabilistic Boolean networks}\label{Boolean}
Boolean networks were introduced by Kauffman~\cite{Kau} to model gene regulatory networks. They have since been applied to other biological processes and have also been studied in other areas such as engineering and computer science (see~\cite{Val} for some recent examples). In general, Boolean networks can be used to model networks where nodes influence the activity of adjacent nodes. The nodes take Boolean values and the network evolves according to Boolean functions assigned with each node. On the other hand, probabilistic Boolean networks (PBNs) were introduced by Shmulevich et al.~\cite{Shm} to account for uncertainties in gene regulatory networks. In PBNs, the Boolean function for each node is chosen from a set of Boolean functions according to a specified probability distribution. 

There are a variety of methods for analyzing Boolean networks and PBNs~\cite{Aku}, among which is the semi-tensor product approach developed by Cheng et al.~\cite{Che}. The use of PBNs to study the ASEP was introduced by Zhao and Krishnan~\cite{Zha}, incorporating many of Cheng et al.'s~\cite{Che} techniques. We now briefly recall this PBN approach.

The \emph{state} or \emph{configuration} of the lattice at time $t$ is represented by a vector 
\[\boldsymbol{x}(t) = (x_1(t),x_2(t),\ldots,x_N(t))\,,\]
where $x_i(t)=0$ if the $i$-th position is empty and $x_i(t)=1$ if it is occupied. For instance, the state in Figure~\ref{transallow} is represented by the vector $(1,0,0,1,0,0,1,0,0,1)$. We will drop the time parameter and simply write $\boldsymbol{x}$ when speaking of states in general.

A transition $\tau$ is described by the system of equations
\begin{align*}
	\tau(\boldsymbol{x}(t)) &= \boldsymbol{x}(t+1)
	\begin{cases}
	x_1(t+1) &= f_1(x_1(t),x_2(t),\ldots,x_N(t))\\
	x_2(t+1) &= f_2(x_1(t),x_2(t),\ldots,x_N(t))\\
	&~\vdots \\
	x_N(t+1) &= f_N(x_1(t),x_2(t),\ldots,x_N(t))\,,
	\end{cases}
\end{align*}
where each $f_i$ is a Boolean function over the state space $\{0,1\}^N$. 

The Boolean functions for some classical ASEP transitions are as follows. For particle entry, these are given by
\begin{align*}
x_i(t+1) &=\begin{cases}
\neg x_i(t) \vee x_i(t), &i=1\\
x_i(t+1), &i\neq 1\,,
\end{cases}
\end{align*}
where $\wedge$, $\vee$ and $\neg$ denote conjunction, disjunction and negation, respectively. For particle exit, the Boolean functions are given by
\begin{align*}
	x_i(t+1) &=\begin{cases}
		\neg x_i(t) \wedge x_i(t), &i=N\\
		x_i(t), &i\neq N\,.
	\end{cases}
\end{align*}
Meanwhile, the Boolean functions for a particle hopping from site $j$ to $j+1$ are as follows
\begin{align*}
	x_i(t+1) &=\begin{cases}
		x_{i}(t) \wedge \bigl(\neg x_j(t) \vee x_{j+1}(t)\bigr), &i=j\\
		x_{i}(t) \vee \bigl(x_j(t) \wedge \neg x_{j+1}(t)\bigr), &i=j+1\\
		x_i(t), &\mbox{otherwise}\,. 
	\end{cases}
\end{align*}
 
The \emph{transition matrix} $M$ is the matrix with entries $M_{r,s}=P(\boldsymbol{x}^{(r)}\to \boldsymbol{x}^{(s)})$, where ${P\left(\boldsymbol{x}^{(r)}\to \boldsymbol{x}^{(s)}\right)}$ is the transition probability from state $\boldsymbol{x}^{(r)}$ to state $\boldsymbol{x}^{(s)}$.

The \emph{steady state distribution} is given by the vector $\pi$ satisfying $M\pi=\pi$, or equivalently,
\[\lim\limits_{k\to\infty} M^k = [\pi~\pi~\ldots~\pi]\,.\]
The $r$-th entry of $\pi$ then gives the \emph{steady state probability} of $\boldsymbol{x}^{(r)}$.

The PBN approach by Zhao and Krishnan~\cite{Zha} provides a method for computing $M$, and hence, $\pi$. Other steady state profiles such as the state density, codon density and translation can also be obtained from $M$. The transition matrix is computed as follows. Let $\mathcal T$ be the set of transitions and $\tau\in\mathcal T$. For each $i=1,2,\ldots,N$, we find a Boolean function $f_{i}^{\tau}$ such that $f_i^{\tau}(\boldsymbol{x}(t))=x_i(t+1)$ under the transition $\tau$. Once $f_i^{\tau}$ is determined, we use semi-tensor product techniques to find the \emph{structure matrix} for $f_i^{\tau}$, that is, the matrix $M_{f_i^{\tau}}$ such
\begin{align}
\delta^{2-f_i^{\tau}(\boldsymbol{x}(t))}_2=M_{f_i^{\tau}}\ltimes_{i=1}^n \delta^{2-x_i(t)}_{2}\,, \label{leftsemi}
\end{align}
where $\delta_{j}^{k}$ is the $j\times 1$ column vector with an entry of $1$ in row $k$ and is zero elsewhere, and $\ltimes$ denotes the semi-tensor product. (See the Appendix for the definitions of specialized matrix products.)

Identity~\eqref{leftsemi} is usually written as $f_i^{\tau}(\boldsymbol{x}(t)) = M_{f_i^{\tau}}\ltimes_{i=1}^n x_i(t)$, with the association $~1\sim\delta^1_2$ and $0\sim\delta^2_2$ implicitly assumed. On the other hand, the \emph{structure matrix} for $\tau$ is the matrix satisfying
\begin{align}
\tau(\boldsymbol{x}(t)) = M^{\tau}\ltimes_{i=1}^N \delta^{2-x_i(t)}_{2}\,, \label{structmattau}
\end{align}
Zhao and Krishnan~\cite{Zha} showed that $M^\tau$ is given by
\[
M^{\tau} = \ast_{i=1}^N M_{f^{\tau}_i}
\]
where $\ast$ denotes the Khatri-Rao product, and that the transition matrix is computed as
\begin{align}
	M=\sum\limits_{\tau\in\mathcal T} p(\tau) M^{\tau}\,, \label{sum}
\end{align}
where let $p({\tau})$ is the transition probability for $\tau$.  The method is summarized in Algorithm~\ref{tasepalg}.

\begin{algorithm}[h!]
	\caption{Computing the steady state distribution}\label{tasepalg}
	\textbf{Input:} Lattice length $N$, Boolean functions $\{f^{\tau}_i\}_{\substack{\tau\in\mathcal{T} \\ 1\leq i\leq N}}$, transition rates $\{r(\tau)\}_{\tau\in\mathcal T}$
	
	\textbf{Output:} The steady state distribution $\pi$\
	
	\For{each $\tau\in\mathcal{T}$}{
		\For{each $i=1,2,\ldots,N$}{
			Compute $M^{\tau}_i$ from $f^{\tau}_i$;
		}
	$M^{\tau} = \ast_{i=1}^n M^{\tau}_i$\;
	}		
	$M = \sum\limits_{\tau\in\mathcal T} \rho(\tau) M^{\tau}$\;
	\KwRet{$\pi$, \mbox{where} $\lim\limits_{k\to\infty} M^k = [\pi~\pi~\cdots~\pi]$}\;
\end{algorithm}

Note that the transition probabilities $p(\tau)$ are not necessarily the transition rates $\alpha$, $\beta$, etc. but rather the \emph{normalized} values of the transition rates. To be more precise, if the transition rate of $\tau$ is $r(\tau)$, then $p(\tau)=r(\tau)/\left(\sum_{\tau'\in\mathcal T} r(\tau')\right)$. As an example, in the TASEP with $N=3$, the transitions are entry, exit, hopping from site $1$ to $2$ and hopping from site $2$ to $3$. The transition probabilities are, respectively, $\alpha/\sigma$, $\beta/\sigma$, $p/\sigma$ and $p/\sigma$, where $\sigma=\alpha+\beta+2p$.

In certain cases $\pi$ can be computed more efficiently by restricting the transition matrix only to the set of allowable states. That is, if $E\subseteq\{0,1\}^N$ is the set of states that the lattice can assume, then the steady state distribution $\pi$ is given by $M|_E\,\pi = \pi$. Take for example Zhao and Krishnan's~\cite{Zha} model where ribosomes travel in groups of $r$. If $r=2$ and $n=5$, then $(1,0,0,0,0)\notin E$. To determine which rows and columns of $M$ to remove, we associate with each state $\boldsymbol{x}$ the $N$-digit binary number whose $i$-th digit from the left is $x_i$ and denote by $\mathrm{Dec}(\boldsymbol{x})$ the decimal value of this binary number. Assume that the states satisfy the obvious lexicographic ordering induced by the alphabet $\{0,1\}$ and denote by $\mathrm{ord}(x)$ the state's lexicographic order. Then, $\mathrm{ord}(\boldsymbol{x})=\mathrm{Dec}(\boldsymbol{x})+1$ gives the index of the row and column corresponding to $\boldsymbol{x}$. In the example where $\boldsymbol{x}=(1,0,0,0,0)$, $\mathrm{Dec}(10000)=16$ and thus, we remove the $17$-th row and column of the transition matrix. For the classical single-species ASEP where the particles occupy one codon at a time, no reduction to the transition matrix is possible since all $2^N$ states are allowable.

\section{Structure matrices of TASEP Boolean functions}\label{matrix}

In the TASEP where entry, exit and unidirectional hopping are the only possible transitions, there is a total of $|\mathcal{T}|=N+1$ transitions and thus, $N(N+1)$ structure matrices of Boolean functions that need to be computed. Expectedly, running Algorithm~\ref{tasepalg} every time a parameter is to be changed can be costly. One way to bypass this algorithm and compute the transition matrix directly using Identity~\eqref{sum} is to determine the explicit forms of the structure matrices, if possible. Fortunately, at least for common transitions in the ASEP, the structure matrices have nice explicit forms. These are enumerated in Table~\ref{struc} together with the structure matrices for \emph{Langmuir kinetics}, where particles can attach or detach at any site of the lattice. These matrices can be proved inductively or constructed using the methods in~\cite[Chapters 2 and 3]{Che}.

Let $\mathbf{1}_k$ and $\mathbf{0}_k$ denote the $k\times k$ identity matrix and $k\times k$ zero matrix, respectively. The structure matrices are expressed in terms of Kronecker products involving the following special matrices.

\begin{align*}
	\mathcal A(N,i) &= 
	\begin{bmatrix}
		\mathbf{1}_{2^{N-i}} & 	\mathbf{1}_{2^{N-i}} \\
		\mathbf{0}_{2^{N-i}} & \mathbf{0}_{2^{N-i}}
	\end{bmatrix} \\
	\mathcal D(N,i) &= 
	\begin{bmatrix}
		\mathbf{0}_{2^{N-i}} & 	\mathbf{0}_{2^{N-i}} \\
		\mathbf{1}_{2^{N-i}} & \mathbf{1}_{2^{N-i}}
	\end{bmatrix} \\
	\mathcal R(N,i) &= 
	\begin{bmatrix}
		\mathbf{1}_{2^{N-i-1}} & \mathbf{0}_{2^{N-i-1}} & \mathbf{0}_{2^{N-i-1}} & \mathbf{0}_{2^{N-i-1}} & \\
		\mathbf{0}_{2^{N-i-1}} & \mathbf{0}_{2^{N-i-1}} & \mathbf{0}_{2^{N-i-1}} & \mathbf{0}_{2^{N-i-1}} & \\
		\mathbf{0}_{2^{N-i-1}} & \mathbf{1}_{2^{N-i-1}} & \mathbf{1}_{2^{N-i-1}} & \mathbf{0}_{2^{N-i-1}} & \\
		\mathbf{0}_{2^{N-i-1}} & \mathbf{0}_{2^{N-i-1}} & \mathbf{0}_{2^{N-i-1}} & \mathbf{1}_{2^{N-i-1}} & \\
	\end{bmatrix} \\
	\mathcal L(N,i) &= 
	\begin{bmatrix}
		\mathbf{1}_{2^{N-i-1}} & \mathbf{0}_{2^{N-i-1}} & \mathbf{0}_{2^{N-i-1}} & \mathbf{0}_{2^{N-i-1}} & \\
		\mathbf{0}_{2^{N-i-1}} & \mathbf{1}_{2^{N-i-1}} & \mathbf{1}_{2^{N-i-1}} & \mathbf{0}_{2^{N-i-1}} & \\
		\mathbf{0}_{2^{N-i-1}} & \mathbf{0}_{2^{N-i-1}} & \mathbf{0}_{2^{N-i-1}} & \mathbf{0}_{2^{N-i-1}} & \\
		\mathbf{0}_{2^{N-i-1}} & \mathbf{0}_{2^{N-i-1}} & \mathbf{0}_{2^{N-i-1}} & \mathbf{1}_{2^{N-i-1}} & \\
	\end{bmatrix}
\end{align*}

\begin{table}[h!]
	\centering
	\caption{Structure matrices in the ASEP with Langmuir Kinetics}
	\label{struc}
	\begin{tabular}{l l}
		\toprule Transition & Structure Matrix \\
		\hline Attaching to site $i$ \vphantom{$\Biggl.\Biggr.$} & 
		
		$\mathbf{1}_{2^{i-1}} \otimes \mathcal A(N,i)$ \\
		Left entry & $[1] \otimes \mathcal A(N,1)$ \\
		Right entry & $\mathbf{1}_{2^{N-1}} \otimes 
		\begin{bmatrix}
			1 & 1 \\
			0 & 0
		\end{bmatrix}$ \\
		Detaching from site $i$ \vphantom{$\Biggl.\Biggr.$} & 
		$\mathbf{1}_{2^{i-1}} \otimes  \mathcal{D}(N,i)$\\
		Left exit & $[1] \otimes \mathcal{D}(N,1)$ \\
		Right exit & $\mathbf{1}_{2^{N-1}} \otimes 
		\begin{bmatrix}
			0 & 0 \\
			1 & 1
		\end{bmatrix}$ \\
		Hopping from site $i$ to $i+1$, $i=1,2,\ldots,N-1$ & $\mathbf{1}_{2^{i-1}} \otimes \mathcal{R}(N,i)$ \\
		Hopping from site $i$ to $i-1$, $i=2,\ldots,N$ & $\mathbf{1}_{2^{i-1}} \otimes \mathcal{L}(N,i)$ \\
		\bottomrule
	\end{tabular}
\end{table}

To give a minimal example, consider the TASEP where $N=2$. The transitions are entry, exit and hopping from site 1 to 2. Their structure matrices are given, respectively, as follows
\[
M_1=\begin{bmatrix}
	1 & 0 & 1 & 0 \\
	0 & 1 & 0 & 1 \\
	0 & 0 & 0 & 0 \\
	0 & 0 & 0 & 0	
\end{bmatrix}, 
M_2=\begin{bmatrix}
	0 & 0 & 0 & 0 \\
	1 & 1 & 0 & 0 \\
	0 & 0 & 0 & 0 \\
	0 & 0 & 1 & 1	
\end{bmatrix}\mbox{ and }
M_3\begin{bmatrix}
	1 & 0 & 0 & 0 \\
	0 & 0 & 0 & 0 \\
	0 & 1 & 1 & 0 \\
	0 & 0 & 0 & 1	
\end{bmatrix}\,.
\]
If $\alpha=0.2$, $\beta=0.3$ and $p=0.5$, then using Identity~\eqref{sum} we obtain the transition matrix  
\[
M=\alpha M_1 + \beta M_2 + p M_3 =\begin{bmatrix}
	0.7 & 0   & 0.2 & 0   \\
	0.3 & 0.5 &	0   & 0.2 \\
	0   & 0.5 & 0.5 & 0   \\
	0   & 0   & 0.3 & 0.8
\end{bmatrix}\,.	
\]

This matrix leads to the steady state distribution $[0.16~0.24~0.24~0.36]$. From left to right, the entries of this vector are the steady state probabilities of the states $(0,0)$, $(0,1)$, $(1,0)$ and $(1,1)$, respectively.

\section{A simplified method for writing the structure matrices}\label{simplified}

A key step in Algorithm~\ref{tasepalg} is writing the structure matrix $M_{f^{\tau}_i}$  for each transition $\tau$ at site $i$, given by the $N$-ary Boolean function $x_i(t+1)=f_i(x_1(t),x_2(t),\ldots,x_N(t))$, where each $f_i$ is expressed in terms of the logical operators $\wedge$, $\vee$ and $\neg$. For an arbitrary Boolean function $f$, let us write its structure matrix by $M_f$. In this section, we give a simplified method for determining $M_f$. 

In order to appreciate the advantage of this method, we first outline the known method, as discussed in~\cite[Chapters 2 and 3]{Che}. Each of the logical operator $\wedge$, $\vee$ and $\neg$ have their corresponding structure matrices given, respectively, by $M_c=\begin{bmatrix} 1 & 0 & 0 & 0 \\ 0 & 1 & 1 & 1 \end{bmatrix}$, $M_d=\begin{bmatrix} 1 & 1 & 1 & 0 \\ 0 & 0 & 0 & 1 \end{bmatrix}$ and $M_n=\begin{bmatrix} 0 & 1 \\ 1 & 0 \end{bmatrix}$. The  idea is to first rewrite the Boolean function $f$ as a semi-tensor product of the structure matrices and the arguments $x_i$, which in this case are represented by $\delta_2^2$ if $x_i=0$ and $\delta_1^2$ if $x_i=1$. For instance, the Boolean function (see~\cite[Example 3.1]{Che}) $f(p,q,r)=(p\wedge \neg q) \vee (r\wedge p)$ is rewritten as \[f(p,q,r)=M_d \ltimes M_c \ltimes p \ltimes M_n \ltimes q \ltimes M_c \ltimes r \ltimes p\,.\]

The next step, which can be quite involved, is to algebraically manipulate the semi-tensor product such that it ends with $p \ltimes q \ltimes r$, with $p$, $q$, and $r$ not occurring elsewhere. This is done using variable swap matrices $W$ and the power-reducing matrix $M_r$. In the current example, one eventually obtains \[f(p,q,r)=M_d \ltimes M_c \ltimes (I_2 \otimes M_n) \ltimes (I_4 \otimes M_c) \ltimes (I_2 \otimes W) \ltimes M_r \ltimes p \ltimes q \ltimes r\,.\] Finally, the structure matrix of $f$ is given by 
\[M_f=M_d \ltimes M_c \ltimes (I_2 \otimes M_n) \ltimes (I_4 \otimes M_c) \ltimes (I_2 \otimes W) \ltimes M_r\,.\]
This method has also been generalized for multi-valued logic.

In order to compute $M_f$ as a semi-tensor product, the Boolean function $f$ must be expressed in terms of logical operators. This may be a challenge for more complicated transitions, especially in the case of multi-valued logic networks. The method we introduce in this section does not use this requirement in the sense that the functions $f_i$ can be expressed in any form.

\subsection{The Boolean case}

For $1\leq k \leq 2^N$, denote by $\mathrm{Bin}_N(k)$ the $N$-digit binary representation of the decimal number $k$. Recall that $\delta_j^k$ is the $j\times 1$ column vector that is zero everywhere except at row $k$ where it has entry $1$. It is straightforward to show via an inductive argument that if a state $\boldsymbol{x}$ of length $N$ has lexicographic order $j$, that is, $\mathrm{Dec}(\boldsymbol{x})=j-1$, then
\begin{align}
\ltimes_{i=1}^N \delta^{2-x_i}_{2} = \delta_{2^N}^{2^N-\mathrm{Dec}(\boldsymbol{x})}\,. \label{semitensorsimp}
\end{align}
We can therefore associate the following quantities with a state  $\boldsymbol{x}=(x_1,x_2,\ldots,x_N)$ uniquely:
\begin{enumerate}
	\item An integer $1\leq j \leq 2^N$, where $j=\mathrm{ord}(x)$
	\item A column vector $\delta^{2^N-\mathrm{Dec}(\boldsymbol{x})}_{2^N}$
	\item An $N$-digit binary number $\mathrm{Bin}_N(\mathrm{Dec}(\boldsymbol{x}))$.
\end{enumerate}
For example, $(0,1,1)\sim j=4 \sim \delta_{8}^{8-3} \sim 011$. For simplicity, since the state and its binary number representation are similar, we will not distinguish between them when evaluating $f$, e.g., $f((0,1,1))=f(011)$.

The theorem that follows gives our result for the Boolean case.

\begin{theorem}\label{thm1} Let $f(x_1,x_2\ldots,x_N)$ be a Boolean function. Then,  for $1\leq j\leq 2^N$,  the $j$-th column of its structure matrix $M_f$ is given by the vector 
	\[
	\begin{bmatrix}
		f(\mathrm{Bin}_N(2^N-j)) \\
		\neg f(\mathrm{Bin}_N(2^N-j))
	\end{bmatrix}\,.
	\]
	That is,
	\[
	M_f = \delta_2 \left[\left(2-f(\mathrm{Bin}_N(2^N-1))\right)~\left(2-f(\mathrm{Bin}_N(2^N-2))\right)~\cdots~\left(2-f(\mathrm{Bin}_N(0))\right)\right]
	\]
	
\begin{proof}Let $1\leq j \leq 2^N$. Then, there exists a unique state $\boldsymbol x$ such that $j=2^N-\mathrm{Dec}(\boldsymbol x)$. Next, let $f$ be a Boolean function and $M_f$ its structure matrix. Since $f(\boldsymbol x)=0$ or $1$, $M_f \ltimes_{i=1}^N \delta_2^{2-x_i}=\delta_2^2$ or $\delta_2^1$. In particular, $M_f \ltimes_{i=1}^N \delta_2^{2-x_i} \sim \begin{bmatrix} f(\boldsymbol x) \\ \neg f(\boldsymbol x) \end{bmatrix}$. Since $\boldsymbol x\sim \delta_{2^N}^{2^N-\mathrm{Dec}(x)}$ by Identity~\eqref{semitensorsimp}, $\begin{bmatrix} f( \boldsymbol x) \\ \neg f(\boldsymbol x) \end{bmatrix}$ is also the $j$-th column of $M_f$, where $j=2^N-\mathrm{Dec}(\boldsymbol x)$. Now, 
	\begin{align*}
		\mathrm{Dec}(\boldsymbol x) &= 2^N -j \\
		\mathrm{Bin}_N (\mathrm{Dec}(\boldsymbol x)) &= \mathrm{Bin}_N (2^N-j) \\
		\boldsymbol x &= \mathrm{Bin}_N (2^N-j)\\
		f(\boldsymbol x) &= f(\mathrm{Bin}_N (2^N-j))\,.
	\end{align*}
\end{proof}	
\end{theorem}

Consider again the Boolean function $f(p,q,r)=(p\wedge \neg q) \vee (r\wedge p)$ in the earlier example. Suppose we wish to determine the fourth ($j=4$) column of $M_f$. The state with this lexicographic order is $\boldsymbol x=\mathrm{Bin}_3(2^3-4)=100$. Then, $f(100)=1$ so that the fourth column of $M_f$ is given by $\delta_{2}^1$. Doing this for all the other columns, that is, for $i=1,2,\ldots,8$, we obtain
\[
M_f = \begin{bmatrix} 
	1 & 0 & 1 & 1 & 0 & 0 & 0 & 0 \\
	0 & 1 & 0 & 0 & 1 & 1 & 1 & 1
\end{bmatrix}\,.
\]

\subsection{The multi-valued logic case}

In $m$-valued logic, variables take values from the set ${\mathcal D_m=\{0,1,2,\ldots,m-1\}}.$ A $N$-ary $m$-valued logical function is a mapping $f:\mathcal{D}_m^N \to \mathcal{D}_m$. We take $m$ to be the positive integer so that the system has $m-1$ species or types of particles. We identify with each $i\in\mathcal{D}_m$ the vector $\delta_m^{m-i}$ and denote the matrix whose $j$-th row is $\delta_m^{d_j}$ by
\[
\delta_m [d_1~d_2~\cdots~d_j~\cdots]\,.
\]

The \emph{structure matrix} $M_f$ of $f$ satisfies 
\begin{align*}
	\delta^{m-f(\boldsymbol{x})}_m=M_f\ltimes_{i=1}^N \delta^{m-x_i}_{m}\,,
\end{align*}
for every state $\boldsymbol{x}\in \mathcal D^N_m$.

Let $\mathrm{B}_{N,m}(n)$ denote the $N$-digit representation in base $m$ of a decimal number $n$. Given a state $\boldsymbol{x}$, denote by $\mathrm{ord}(\boldsymbol{x})$ its lexicographic order and $\mathrm{Dec}_m(\boldsymbol{x})$ the decimal value of the base $m$ number whose digits are the $x_i$'s. As in the Boolean case, $\mathrm{ord}(\boldsymbol{x})=\mathrm{Dec}_m(\boldsymbol{x})+1$. Similar to Identity~\ref{semitensorsimp}, we can show that the following identity holds.
\begin{align*}
\ltimes_{i=1}^N \delta_m^{m-x_i} = \delta_{m^N}^{m^N-\mathrm{Dec}_m(\boldsymbol{x})}\,.
\end{align*}

For each state $\boldsymbol{x}$, we can therefore associate the following quantities uniquely:
\begin{enumerate}
	\item An integer $1\leq j \leq m^N$, where $j=\mathrm{ord}(\boldsymbol{x})$
	\item A column vector $\delta^{m^N-\mathrm{Dec}_m(\boldsymbol{x})}_{m^N}$
	\item An $N$-digit base $m$ number $\mathrm{B}_N(\mathrm{Dec}_m(\boldsymbol{x}))$.
\end{enumerate}
For example, if $m=3$, then $(0,1,0,2)\sim j=12 \sim \delta_{81}^{81-11} \sim 0102$. Similar to the Boolean case, we take a state and its $N$-digit base $m$ numerical representation as identical when evaluating $m$-valued logical functions, e.g., $f((0,1,0,2))=f(0102)$.

The next theorem gives the generalization of Theorem~\ref{thm1} to $m$-valued logic. The proof is completely analogous and is therefore omitted.

\begin{theorem}\label{thm2} Let $f(x_1,x_2\ldots,x_N)$ be an $N$-ary $m$-valued logical function. Then, for $1\leq j\leq m^N$, the $j$-th column of its structure matrix $M_f$ is given by the vector $\delta_{m}^{m-f(\mathrm{B}_{N,m}(m^N-j))}$. In other words,
	\[
	M_f = \delta_m\left[\left(m-f(\mathrm{B}_{N,m}(m^N-1))\right)~\left(m-f(\mathrm{B}_{N,m}(m^N-2))\right)~\cdots~\left(m-f(\mathrm{B}_{N,m}(0))\right)\right]\,.
	\]
\end{theorem}

For example, let $N=2$ and $m=3$ and consider the transition $\tau$ where a particle of type $2$ enters the leftmost site of the lattice. Then, the transition is described by the following system of equations.
\begin{align*}
\tau(\boldsymbol{x}(t)) &= \begin{cases}
x_1(t+1) &= f_1(x_1,x_2) = 
\begin{cases}
	x_1, &x_1\neq 0\\
	2, &x_1 =0	
\end{cases}  \\
x_2(t+1) &= f_2(x_1,x_2) = x_2\,.	
\end{cases}	
\end{align*}
Note that while it is possible to write $f_1$ in terms of $m$-valued logical operators, this is not necessary in order to apply Theorem~\ref{thm2}. The first column of $M_{f_1}$ is then given by the vector
\[
\delta_{3}^{3-f(B_{2,3}(8))} = \delta_{3}^{3-f(22)} = \delta_{3}^1\,.
\]
Computing the other columns gives us
\[
M_{f_1} = \begin{bmatrix}
 1 & 1 & 1 & 0 & 0 & 0 & 1 & 1 & 1 \\
 0 & 0 & 0 & 1 & 1 & 1 & 0 & 0 & 0 \\
 0 & 0 & 0 & 0 & 0 & 0 & 0 & 0 & 0
\end{bmatrix}\,.
\]
For $f_2$, the structure matrix is given by
\[
M_{f_2} = \begin{bmatrix}
	1 & 1 & 1 & 0 & 0 & 0 & 0 & 0 & 0 \\
	0 & 0 & 0 & 1 & 1 & 1 & 0 & 0 & 0 \\
	0 & 0 & 0 & 0 & 0 & 0 & 1 & 1 & 1
\end{bmatrix}\,.
\]
Using Identity~\eqref{structmattau}, the structure matrix for the entire transition is determined to be
\[
M^{\tau} = M_{f_1} * M_{f_2} = \delta_9[1~2~3~4~5~6~1~2~3]\,.
\]

\section{A general method for writing Boolean functions}\label{general}

A key advantage of the PBN approach is its ability to accommodate other transitions as long as the corresponding logical functions can be written explicitly. In this section, we describe methods for writing these functions in a systematic manner.

\subsection{The Boolean case}

Suppose that a transition involves site $j$ changing from $0$ into $1$. Let $\mathbb I_0$ (respectively, $\mathbb I_1$) be the set of indices such that for every $i\in\mathbb I_0$, $x_i(t)$ must be 0 (respectively, 1) for the transition to occur, otherwise the state does not change in the next time step. Then,
\begin{align}
x_j(t+1) = x_j(t) \vee \left[ \left(\neg \bigvee_{i\in\mathbb I_0} x_i(t)\right) \wedge\left( \bigwedge_{i\in\mathbb I_1 } x_i(t)\right) \right]\,. \label{0to1}
\end{align}
If for some $i_0\in\mathbb I_0$, $x_{i_0}(t)=1$ or for some $i_1\in\mathbb I_1$, $x_{i_1}(t)=0$, then the entire expression inside square brackets reduces to $0$. In which case, $x_j(t+1)=x_j(t) \vee 0 = x_j(t)$ so that the site does not change, as desired. Note that necessarily, $j\in I_0$. On the other hand, if for every $i_0\in\mathbb I_0$, $x_{i_1}(t)=0$ and for every $i_1\in\mathbb I_1$, $x_{i_1}(t)=1$, then expression inside square brackets becomes $1$ and  $x_j(t+1)=x_j(t)\vee 1 = 1$.

Meanwhile, if a transition involves site $j$ changing from $1$ into $0$, with $\mathbb I_0$ and $\mathbb I_1$ defined as before, we have
\begin{align}
x_j(t+1) = x_j(t) \wedge \left[ \left(\bigvee_{i\in\mathbb I_0} x_i(t)\right) \vee \left( \neg \bigwedge_{i\in\mathbb I_1 }  x_i(t)\right) \right]\,. \label{1to0}
\end{align}
Using a similar analysis as in the previous case, we can show that this Boolean function guarantees that $x_j(t+1)=0$ if all conditions are satisfied and $x_j(t+1)=x_j(t)$, otherwise.

In addition to classical transitions such as entry, exit and hopping, the Boolean functions for the following dynamics can also be written using Identities~\eqref{0to1}~and~\eqref{1to0}. We leave this task to the reader.
\begin{enumerate} 
	\item \emph{Parallel memory reservoir}. Particles in the main lattice may hop to and from a parallel site in the reservoir lattice, but the particles in the reservoir lattice may not hop sidewards~\cite{Eza}.
	\item \emph{Parallel lattices}. Particles travel across multiple one-dimensional lattices. The lattices may or may not interact with each other.
	\item \emph{Overtaking}. Particles may overtake other species of particles in one or both directions.
	\item \emph{Long-range hopping}. Particles may hop according to a specified jump length $l$.
	\item \emph{Periodic boundary}. Particles may hop from one endpoint of the lattice to another, subject to the prescribed direction.
\end{enumerate}

Section~\ref{multifcn} shows a method for writing the logical functions in the multi-valued logic case. Alternatively, we can also ``Booleanize'' a multi-valued state so that the Boolean method described here can be used. To do this, we represent each state in a system with $m-1$ species of particles by an $m-1$ by $N$ array $\boldsymbol{x}=x_{n,i}$, where $x_{n,i}=1$ if a particle of type $n$ occupies position $i$ and $0$ if otherwise. For example, the array
\[\boldsymbol{x}=\begin{bmatrix}
	0 & 0 & 0 & 0 & 0 \\
	1 & 0 & 0 & 0 & 0 \\
	0 & 1 & 0 & 0 & 0 \\
	0 & 0 & 0 & 1 & 0
\end{bmatrix}\]
corresponds to the multi-valued state $(2,3,0,4,0)$. This representation results into a transition matrix containing rows and columns corresponding to states that do not exist, particularly arrays with more than one $1$ in a column, in addition to states that are not allowed by the transitions and other restrictions in the system. As before, to compute the steady state distribution more efficiently, the size of the resulting transition matrix can be reduced by restricting its indices to allowable states.

Using this method, we can write some of the Boolean functions for a multi-particle system as follows. The attachment of a particle of type $m$ at site $j$ is given by
	\[
	x_{n,i}(t+1) =
	\begin{cases}
	x_{n,i}(t) \vee \neg \left( \bigwedge\limits_{l\neq m} x_{l,j}(t)\right), &(n,i)=(m,j)\\
	x_{n,i}(t), &\mbox{otherwise}\,.	
	\end{cases}
	\]
the detachment of a particle of type $m$ at site $j$ is given by
	\[
	x_{n,i}(t+1) =
	\begin{cases}
		x_{n,i}(t) \wedge \left(\bigvee\limits_{l\neq m} x_{l,j}(t)\right), &(n,i)=(m,j)\\
		x_{n,i}(t), &\mbox{otherwise}\,.	
	\end{cases}
	\]
For the hopping of a particle of type $m$ at site $j$ to the right at site $j+1$, we have
\begin{align*}
	x_{n,i}(t+1) =
\begin{cases}
	x_{n,i}(t) \wedge \left[ \left(\bigvee\limits_{l\neq m} x_{l,j}(t)\right) \vee \left(\bigvee\limits_{l=1}^N x_{l,j+1}(t)\right) \right], &(n,i)=(m,j)\\
	x_{n,i}(t) \vee \left[ \left(\neg \bigvee\limits_{l\neq m} x_{l,j}(t) \right) \wedge \left(\neg \bigvee\limits_{l=1}^N x_{l,j+1}(t)\right) \wedge x_{m,j}(t)\right], &(n,i)=(m,j+1)\\
	x_{n,i}(t), &\mbox{otherwise}\,.
\end{cases}	
\end{align*}

\subsection{The multi-valued logic case} \label{multifcn}

We now describe a general method for writing the $m$-valued logical functions that describe transitions in a multi-species system. This method does not use any unary or binary multi-valued logical operators other than conjunction and disjunction. Instead, it uses the function $\sigma_{a,b}(P)$ which can be implemented efficiently in code.

In the multi-valued logic setting, conjunction and disjunction are given by
\begin{align*}
	x_1 \wedge x_2 &= \min (x_1,x_2) \\
	x_1 \vee x_2 &= \max (x_1,x_2)\,.
\end{align*}

Let $P={p_1,p_2,\ldots,p_{|P|}}$ be a set of Boolean-valued statements and define the function $\sigma_{a,b} (P)$ as follows
\[
\sigma_{a,b} (P) = \begin{cases}
a, &p_1=p_2=\cdots=p_{|P|}=1\\
b, &\mbox{otherwise}\,.	
\end{cases}
\]
In other words, $\sigma_{a,b}(P)$ is equal to $1$ if all statements in the set $P$ are true, and $0$, otherwise.

Suppose that the system has $m-1$ species of particles. Given a transition $\tau$, let $\mathbb I_{\tau}$ be the set of statements ``$x_i(t)=c$'' all of which need to be satisfied for the transition to happen. Alternatively, we can represent $\mathbb I_{\tau}$ as the collection of ordered pairs $(i,c)$. If a transition $\tau$ involves site $j$ changing from $a$ to $b$, then necessarily, $(j,a)\in \mathbb I_{\tau}$. Furthermore, if $a>b$, then
\[
x_j(t+1) = x_j(t) \wedge \left(\sigma_{b,m-1} \left(\mathbb I_{\tau}\right)\right)\,.
\]
If there exists $(i,c)\in\mathbb I_{\tau}$ such that $x_i(t)\neq c$, then $\sigma_{b,m-1} \left(\mathbb I_{\tau}\right)=m-1$ and thus, $x_j(t+1)=x_j(t) \wedge (m-1) = x_j(t)$. That is, site $j$ remains unchanged. Otherwise, $\sigma_{b,m-1} \left(\mathbb I_{\tau}\right)=b$ and since $a>b$,  $x_j(t+1)=x_j(t) \wedge b = b$.

On the other hand, if $\tau$ involves site $j$ changing from $a$ to $b$ with $a<b$, we have
\[
x_j(t+1) = x_j(t) \vee \left(\sigma_{b,0} \left(\mathbb I_{\tau}\right)\right)\,.
\]

As an example, the attachment of a particle of type $m'$ to site $j$ is given by
\[
x_i(t+1) = 
\begin{cases}
x_i(t) \vee \sigma_{m',0} \left((j,0)\right), &i=j\\
x_i(t), &i\neq j\,,	
\end{cases}\]
while the detachment of a particle of type $m'$ from site $j$ is given by
\[
x_i(t+1) = 
\begin{cases}
	 x_i(t) \wedge \sigma_{0,m-1} \left((j,m')\right), &i=j\\
	x_i(t), &i\neq j\,.
\end{cases}\]
We can also write the $m$-valued functions for other slightly more complicated transitions. For example, the switching of positions between a particle of type $m_1$ at site $j$ with a particle of type $m_2$ at site $j+1$, where $m_1<m_2$, is given by the following pair of $m$-valued logical functions
\begin{align*}
x_i(t+1) &= 
\begin{cases}
	x_i(t) \vee \sigma_{m_2,0} \left((j,m_1),(j+1,m_2)\right), &i=j\\
	x_{i}(t) \wedge \sigma_{m_1,m-1} \left((j,m_1),(j+1,m_2)\right), &i=j+1\\
	x_i(t), &\mbox{else}\,.
\end{cases}	
\end{align*}

\subsection{An illustration}

Consider a TASEP model with $m-1=5$ species of particles and $N=5$ sites. Particles of type $i$, $i=1,2,\ldots,5$ have entry and exit rates given by $\alpha_i=\beta_i=50i/3$. All particles, regardless of type, hop along the lattice from left to right at rate $p=1$ and the system has no other transitions. The two-species version of this model was studied by Bonnin et al~\cite{bon}. In this example, we increased the number of species of particles for illustration purposes.

Suppose that we are interested in determining the effect of the varying entry and exit rates among the different species of particles to the \emph{average density} $\rho_{i,j}$ of a particle of type $i$ at site $j$. If the steady state distribution is given by $\pi$, then 
\[
\rho_{i,j} = \sum_{\boldsymbol{x} \in \mathcal{D}^N_m} \left[x_j=i\right]  \pi_{\mathrm{ord}(\boldsymbol{x})}\,,
\]
where $[\cdot]$ is the Iverson bracket.

Using Algorithm~\ref{tasepalg} and the method for writing $m$-valued logical functions and their structure matrices in Sections~\ref{simplified}~and~\ref{general}, the matrix $R=[\rho_{i,j}]$ is determined to be
\[
R = \begin{bmatrix}
	0.06657086 & 0.04285310 & 0.03331269 & 0.02375500 & 0.001425016 \\
	0.13314175 & 0.08570614 & 0.06662543 & 0.04751040 & 0.001425174 \\ 
	0.19971275 & 0.12855962 & 0.09993798 & 0.07126531 & 0.001424970 \\
	0.26628367 & 0.17141283 & 0.13325085 & 0.09502100 & 0.001424768 \\
	0.33286125 & 0.21433388 & 0.16689607 & 0.11966211 & 0.001435504
\end{bmatrix}\,.
\]

\begin{figure}[h!]
	\centering
	\includegraphics[scale=0.6]{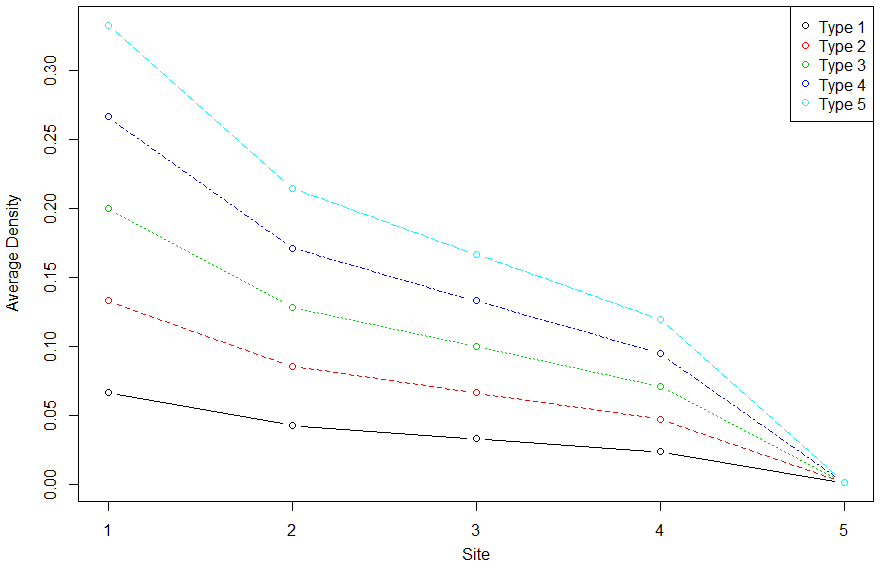}
	\caption{Average densities per site in a system with $5$ species of particles, with $\alpha_i=\beta_i=50i/3$ and $q=1$.}
	\label{averho}
\end{figure}

A graph of the average densities is shown in the Figure~\ref{averho}. We see that the species with the fastest entry rate has the highest average density per site, but the difference is reduced towards the end of the lattice, where all the species of particles have nearly identical average densities.

\section*{Appendix}

Let $X$ and $Y$ be matrices with dimensions $r_1\times c_1$ and $r_2 \times c_2$. Denote the ordinary matrix product by $\times$. The following are the specialized matrix products used in this paper.

\begin{enumerate}
	\item \emph{Kronecker product}
	\[
	X \otimes Y = 
	\begin{bmatrix}
		x_{11}Y & \cdots & x_{1,c_1}Y\\
		\vdots & \ddots & \vdots \\
		x_{r_1,1}Y & \cdots & x_{r_1,c_1}Y\\
	\end{bmatrix}\,.
	\]
	\item (Left) \emph{semi-tensor product}
	\[
	X\ltimes Y = (X\otimes I_{\mathrm{lcm}(c_1,r_2)/c_1}) \times (Y\otimes I_{\mathrm{lcm}(c_1,r_2)/c_1})\,,
	\]
	Observe that the semi-tensor product becomes the ordinary matrix product if $c_1=r_2$.
	\item (Column-wise) \emph{Khatri-Rao product}
	\[	
	X * Y = 
	\begin{bmatrix}
		(x_1 \otimes y_1) \, (x_2 \otimes y_2) \cdots (x_{c_1} \otimes y_{c_2})  
	\end{bmatrix}\,,
	\]
	where $c_1=c_2$, and $x_i$ and $y_i$ are the $i$-th columns of of $X$ and $Y$, respectively.
\end{enumerate}

\nocite{*}
\bibliographystyle{fundam}
\bibliography{citationlist}

\end{document}